\documentstyle[eqsecnum,aps,epsfig]{revtex}
\begin{document}
\draft

      \title{
\begin{flushright} {\small IFT-P.  \,\, gr-qc/0002010} \end{flushright}
             Weak decay of uniformly accelerated protons and related processes
            }


\author{ Daniel A. T. Vanzella and George E. A. Matsas}
\address{Instituto de F\'\i sica Te\'orica, 
         Universidade Estadual Paulista\\
         Rua Pamplona 145\\
         01405-900, S\~ao Paulo, SP\\
         Brazil}
\def\baselinestretch{1.5}
\maketitle


\begin{abstract}

We investigate the weak interaction emission of spin-$1/2$ fermions  from 
accelerated currents. As particular applications, we analyze the
decay of uniformly accelerated protons and neutrons, and 
the neutrino-antineutrino emission from uniformly accelerated 
electrons. The possible relevance of our results to astrophysics 
is also discussed.

\end{abstract}



\section{Introduction}
\label{Intro}

We investigate the weak interaction emission of spin-$1/2$ fermions
from classical and semiclassical currents.  We denote by semiclassical
those currents which possess classical trajectories and are endowed with 
quantized inner energy levels. Our results can be used 
to investigate a broad class of processes involving accelerated 
particles provided that they have a well defined world line, 
as indeed verified in many situations of interest. 
In the case where the particle is accelerated by a background
electromagnetic field, such processes could be fully analyzed
quantum-mechanically. As a consequence, any recoil effects  due to 
the fermion emission would be  automatically taken into account~\cite{BLP,BKS}.
For instance, in Ref.~\cite{BLP} a quasiclassical approach to Quantum 
Electrodynamics
was developed to consider $\gamma$-synchrotron radiation from an electron
in a classical background magnetic field. This  approach
is basically characterized by assuming that the electron motion is
quasiclassical. This is always
possible as far as the magnetic field is not very strong, namely
$H \ll H_0$ and $\gamma \gg 1$, where $H_0=4.4 \, \cdot \, 10^{13} $~Gauss 
and $\gamma$
is the  Lorentz factor for the electron. This quasiclassical 
approach applied for neutrino-antineutrino emission is analyzed in detail
in Sec.~6.1 of Ref.~\cite{BKS}. 

Although this is hard to take into 
account the current recoil in our semiclassical approach, the relations 
here obtained, which agrees with the full quantum mechanical treatment
in the proper limit ($\chi \ll 1$, see e.g. Sec.~\ref{Neutrino emission}), 
are easily applicable  when the process involves particle decay and the 
trajectory itself (rather than the underlying dynamical process which 
generates it) is  inferred from the observational data. 
Explicit results for uniformly accelerated currents are exhibited.

As far as we know, the first ones to call attention to the possibility 
that non-inertial protons may decay were 
Ginzburg and Syrovatskii~\cite{GINZ} but only recently Muller~\cite{M} 
obtained the first estimation for the decay rate associated with the process 
$ p \to n \;e^+  \nu_e$ by assuming that all the involved particles are 
scalars. Here, as a particular application of modeling accelerated 
particles by  semiclassical currents, we perform a comprehensive 
(inertial-frame) analysis of the inverse  $\beta$ decay for uniformly 
accelerated protons. We show that under certain astrophysical conditions, 
high-energy protons in strong background magnetic fields should rapidly decay. 

The observation of non-inertial neutrons is less trivial.
Notwithstanding the calculation of the $\beta$ decay rate 
for accelerated neutrons may be of some relevance in situations where they 
are under the influence of ``relatively'' strong background gravitational 
fields and, thus, will be also performed. 

Some features of the $\beta$ 
and inverse  $\beta$ decays for uniformly accelerated nucleons    
will be  discussed in terms of the  Fulling-Davies-Unruh (FDU) effect. 
The FDU  effect asserts that the Minkowski 
vacuum corresponds to a thermal bath with respect  to uniformly accelerated
observers \cite{FD,U}. It is perhaps remarkable that although inertial 
observers associate the inverse $\beta$ decay 
to the channel $ p \to n \;e^+  \nu_e$, coaccelerated observers
associate the {\em same} proton decay event to one of the following channels:
$
p \; e^- \to n \; \nu_e , 
$
$
p \; {\bar \nu}_e \to n \; e^+ 
$ 
or
$
p \; e^- {\bar \nu}_e \to n\; ,
$
where the absorbed $ e^-$ and ${\bar \nu}_e$ are Rindler particles 
present in the FDU thermal bath ``attached'' to the  proton's frame~\cite{MV}. 
The corresponding branching ratios can be also calculated as
a function of the proton acceleration.

Under a certain restriction, we can make our semiclassical current to 
behave as a classical one. This is suitable to investigate
neutrino-antineutrino emission from accelerated electrons. 
This process is of relevance in some astrophysical situations 
as, e.g., in the cooling of neutron stars. Eventually 
we compare our results in the proper limit with the ones  
in the literature obtained by quantizing electrons in a background 
magnetic field \cite{LP}-\cite{people}.  

The paper is  organized as follows: In Sec.~\ref{DNC} we introduce the
semiclassical currents and discuss how they model particle decays. In 
Sec.~\ref{inertial}, we introduce the weak-interaction action and couple 
our current to a spin-$1/2$ fermion-antifermion field. Afterwards we 
calculate  the differential transition probability for currents 
following arbitrary world lines. In Sec.~\ref{accelerated} 
we use the results obtained in the previous section to 
explicitly evaluate the fermion emission rate and radiated power 
for the particular case of a 
uniformly accelerated current. The next two sections, Sec.~\ref{decay} and
Sec.~\ref{Neutrino emission}, are dedicated to analyze in detail
the decay of uniformly accelerated protons and neutrons,
and the neutrino-antineutrino emission from uniformly accelerated 
electrons, respectively. We also comment on the possible
astrophysical relevance of our results. 
We dedicate Sec.~\ref{Discussions} for our final discussions. 
We will use natural units $c = \hbar = k_B = 1$ throughout this 
paper unless stated otherwise.


\section{Semiclassical vector current}
\label{DNC}

Let us consider a particle in a four-dimensional Minkowski spacetime 
covered by the usual inertial coordinates $(t,{\bf x})\in {\rm R}^4$.
Let $x^\mu(\tau)$ be the particle's world line and $\tau$ its
proper time. The classical vector current associated with this 
particle is given by
\begin{equation}
j^\mu (x) = \frac{q \; u^\mu (\tau)}{u^0 (\tau)} \delta^3
({\bf x}-{\bf x}(\tau)),
\label{C}
\end{equation}
where $q$ is a ``small'' coupling constant and 
$u^\mu (\tau) \equiv dx^\mu/d\tau $. The current above is suitable 
to describe a pointlike classical (i.e., with no-inner structure)
particle. Eventually it can be used to describe a fermion $f_1$ and 
antifermion $\bar f_2$ emission from an accelerated particle $p_1$: 
$$
p_1 \to p_1 f_1 \bar f_2 \;\;\; 
$$ 
(e.g., a non-inertial electron emitting a neutrino-antineutrino pair). 
Notwithstanding, this current must be improved in order to allow 
more general processes of the form 
\begin{equation}
p_1 \to p_2 f_1 \bar f_2\;,
\label{p1p2}
\end{equation}
where particle $p_1$ turns into particle $p_2$ with a fermion-antifermion
pair emission (e.g., decay of an accelerated proton into a neutron 
with a positron-neutrino  emission). This is attained by replacing the 
real coupling constant $q$ by an operator-valued function (see e.g. 
Ref.~\cite{BD})
\begin{equation}
\hat q(\tau) = e^{i \hat H_0 \tau} \hat q_0 e^{-i \hat H_0 \tau} \;\;\; .
\label{QC}
\end{equation}
This can be regarded as the usual first-quantization procedure, where 
a classical observable, $q$, is replaced by a self-adjoint operator, 
$\hat q_0$, evolved by the one-parameter group of unitary operators 
$\hat {\rm U} (\tau) = e^{-i \hat H_0 \tau}$. Here
$\hat H_0$ is the {\it proper} Hamiltonian of the system, i.e.,
\begin{equation}
\hat H_0 \; \vert p_j \rangle = M_j \; \vert p_j \rangle\;,\;\; j=1,2\;,
\label{H0}
\end{equation}
where $\vert p_1 \rangle$ and $\vert p_2 \rangle$ are the energy eigenstates
associated with particles $p_1$ and $p_2$, respectively, and $M_1$ and 
$M_2$ are the corresponding {\it rest masses}.
As a result, the classical current (\ref{C}) is replaced by
the semiclassical one 
\begin{equation}
\hat j^\mu (x)=
\frac{\hat q(\tau)\;u^\mu (\tau)}
{u^0 (\tau)} \delta^3 ({\bf x}-{\bf x}(\tau))\;.
\label{CI}
\end{equation}

Calculating the matrix elements 
$j^\mu_{(p_i \to p_j)} \equiv \langle p_j \vert \hat j^\mu \vert p_i \rangle $
associated with $\hat j^\mu $, we have
\begin{equation}
j^\mu_{(p_i \to p_j)} = G_{\rm eff}\; e^{i(M_j-M_i)\tau}\;
\frac{u^\mu (\tau)}{u^0 (\tau)}\delta^3 ({\bf x}-{\bf x}(\tau))\;\;\; ,
\label{JM}
\end{equation}
where 
$G_{\rm eff} \equiv \vert \langle p_2 \vert \hat q_0 \vert p_1 \rangle \vert$
is the effective coupling constant.
Note that we can recover current (\ref{C}) from Eq.~(\ref{JM})
by making $ M_2 = M_1$ and $G_{\rm eff}=q$.

We will assume  that the fermion emission does
not change appreciably the four-velocity of $p_2$ with respect
to $p_1$. We will denominate this assumption ``no-recoil 
condition.'' This is verified as far as the momentum of the
emitted fermions (with respect to the inertial frame instantaneously 
at rest with the current) satisfies $|{\tilde {\bf k}}| \ll M_1,\;M_2$.
In order to be conservative we will impose 
$|{\tilde {\bf k}}|< {\tilde \omega} \ll M_1,\;M_2$.
It will become clear further that the 
typical energy of the emitted  fermions $\tilde \omega$ 
is of the order of the current's proper acceleration $a$. 
Hence our condition above can be recast in the suitable form
$a \ll M_1,\; M_2$. Our results should be accurate as far as this 
condition is verified.
\setcounter{equation}{0}


\section{Fermion-antifermion emission from a semiclassical current}
\label{inertial}

We shall describe the emitted fermions by spinorial fields 
\begin{equation}
\hat \Psi(x)= \sum_{\sigma = \pm } \int d^3 {\bf k}
\left( \hat b_{{\bf k} \sigma} \psi^{(+\omega)}_{{\bf k} \sigma} (x)
     + \hat d^\dagger_{{\bf k} \sigma} \psi^{(-\omega)}_{-{\bf k} -\sigma} (x) 
\right)\;,
\label{FF}
\end{equation}
where $ \hat b_{{\bf k} \sigma} $ and $ \hat d^\dagger_{{\bf k} \sigma} $ 
are annihilation and creation operators of fermions
and antifermions, respectively, with three-momentum ${\bf k}=(k^x,
k^y,k^z)$ and
polarization $\sigma$. We will adopt the notation used in~\cite{MV}.
Energy $\omega$, momentum ${\bf k}$ and mass $m$ are related as usually: 
$\omega=\sqrt{{\bf k}^2+m^2}>0$. 
$ \psi^{(+\omega)}_{{\bf k} \sigma} $ and $ \psi^{(-\omega)}_{{\bf k} \sigma} $
are positive and negative frequency solutions of the Dirac equation
$i\gamma^\mu \partial_\mu \psi^{(\pm \omega)}_{{\bf k} \sigma} 
 - m \psi^{(\pm \omega)}_{{\bf k} \sigma} =0$.
By using the $\gamma^\mu$ matrices in the 
Dirac representation (see, e.g., Ref.~\cite{IZ}), we find
\begin{equation}
\psi^{(\pm \omega)}_{{\bf k} +} (x) =
 \frac{e^{i(\mp \omega t + {\bf k}\cdot{\bf x})}}
{\sqrt{16 \pi^3 \omega(\omega \pm m)}}
\left(
\begin{array}{c}
m \pm \omega \\
0\\
k^z\\
k^x+i k^y
\end{array}
\right) \;\; 
\label{NM1}
\end{equation}
and
\begin{equation}
\psi^{(\pm \omega)}_{{\bf k} -} (x) = 
\frac{e^{i(\mp \omega t + {\bf k}\cdot{\bf x})}}
{\sqrt{16 \pi^3 \omega(\omega \pm m)}}
\left(
\begin{array}{c}
0\\
m \pm \omega \\
k^x-i k^y\\
-k^z
\end{array}
\right) \; .
\label{NM2}
\end{equation} 
We have orthonormalized modes (\ref{NM1})-(\ref{NM2})
according to the inner product~\cite{BD}: 
\begin{equation}
\langle 
\psi^{(\pm \omega)}_{{\bf k} \sigma} , 
\psi^{(\pm \omega')}_{{\bf k}' \sigma'} 
\rangle 
\equiv 
\int_\Sigma d\Sigma_\mu 
\bar \psi^{(\pm \omega)}_{{\bf k} \sigma} \gamma^\mu 
\psi^{(\pm \omega')}_{{\bf k}' \sigma'}
=
\delta^3({\bf k}-{\bf k}') \delta_{\sigma \sigma'} 
\delta_{\pm \omega \; \pm \omega'} \;\;,
\label{IP}
\end{equation}
where $d\Sigma_\mu \equiv n_\mu d\Sigma $ with $n^\mu$ being a 
unit vector orthogonal to $\Sigma$ and pointing to the future,
and $\Sigma$ is an arbitrary spacelike hypersurface. 
We have chosen $t=const$ for 
the hypersurface $\Sigma$. As a consequence,  canonical 
anticommutation relations for fields and conjugate momenta lead to 
the following simple anticommutation relations for 
creation and annihilation operators:
\begin{equation}
\{\hat b_{{\bf k} \sigma},\hat b^\dagger_{{\bf k}' \sigma'}\}=
\{\hat d_{{\bf k} \sigma},\hat d^\dagger_{{\bf k}' \sigma'}\}=
\delta^3({\bf k}-{\bf k}') \; \delta_{\sigma \sigma'} 
\label{ACR}
\end{equation}
and
\begin{equation}
\{\hat b_{{\bf k} \sigma},\hat b_{{\bf k'} \sigma'}\}=
\{\hat d_{{\bf k} \sigma},\hat d_{{\bf k'} \sigma'}\}=
\{\hat b_{{\bf k} \sigma},\hat d_{{\bf k'} \sigma'}\}=
\{\hat b_{{\bf k} \sigma},\hat d^\dagger_{{\bf k'} \sigma'}\}=
0 \;\; .
\end{equation}

Next we minimally couple the spinorial fields $\hat \Psi_1$ and $\hat \Psi_2$,
associated with the emitted fermions $f_1$ and $\bar f_2$, respectively, 
to our general current $\hat j^{\mu}$ according to the weak-interaction
action
\begin{equation}
\hat S_I = \int d^4x \;\hat j_\mu 
           \{\hat{\bar \Psi}_1 \gamma^\mu (c_V-c_A\gamma^5)\hat \Psi_2 +
            \hat{\bar \Psi}_2 \gamma^\mu (c_V-c_A\gamma^5)\hat \Psi_1 \} \;,
\label{S}
\end{equation}
where $c_V$ and $c_A$ will be settled further.

The vacuum transition amplitude for process (\ref{p1p2}) at the
tree level is given by
\begin{equation}
{\cal A}^{\sigma_1 \sigma_2}_{{\bf k}_1 {\bf k}_2} =
\; \langle  p_2 \vert \otimes \langle {f_1}_{{\bf k}_1 \sigma_1} , 
\bar {f_2}_{{\bf k}_2 \sigma_2} \vert \;
\hat S_I \;
\vert 0 \rangle \otimes \vert p_1  \rangle \; .
\label{AMP}
\end{equation}
Note that the second term inside  the parenthesis  
at the right hand side of Eq.~(\ref{S}) vanishes in this case.
By using the field decomposition (\ref{FF}) in Eq.~(\ref{S}), and acting  
$\hat S_I $ in Eq.~(\ref{AMP}), we obtain
\begin{equation}
{\cal A}^{\sigma_1 \sigma_2}_{{\bf k}_1 {\bf k}_2} =
\int d^4x \; j_{\mu}^{(p_1 \to p_2)}
{\bar \psi}^{(+\omega_1)}_{{\bf k}_1 \sigma_1}
\gamma^{\mu}(c_V-c_A\gamma^5)
\psi^{(-\omega_2)}_{-{\bf k}_2 -\sigma_2}\;,
\label{AMPI}
\end{equation}
where
$j_{\mu}^{(p_i \to p_j)}$ and
$\psi^{(\pm \omega_j)}_{{\bf k}_j \sigma_j}$ are obtained from
Eqs.~(\ref{JM}) and (\ref{NM1})-(\ref{NM2}), respectively.

Letting the amplitude~(\ref{AMPI}) in the expression for the differential
transition probability
\begin{equation}
\frac{d{\cal P}^{p_1 \to p_2}}{d^3{\bf k}_1 d^3{\bf k}_2}\;=\;
\sum_{\sigma_1=\pm} \sum_{\sigma_2=\pm} \vert 
{\cal A}^{\sigma_1 \sigma_2}_{{\bf k}_1 {\bf k}_2} \vert^2\;,
\label{dP1}
\end{equation}
we obtain
\begin{equation}
\frac{d{\cal P}^{p_1 \to p_2}}{d^3{\bf k}_1 d^3{\bf k}_2}\;=\;
\int d^4x \int d^4x' \;
{\rm J}^{(p_1 \to p_2)}_{\mu \nu} (x,x')
{\rm G}_{{\bf k}_1 {\bf k}_2}^{\mu \nu} (x,x') \;,
\label{dP2}
\end{equation}
where
\begin{equation}
{\rm J}^{(p_1 \to p_2)}_{\mu \nu} (x,x') 
\equiv 
j_{\mu}^{(p_1 \to p_2)}(x)j_{\nu}^{(p_2 \to p_1)}(x')\;,
\label{J}
\end{equation}
and
\begin{eqnarray}
{\rm G}_{{\bf k}_1 {\bf k}_2}^{\mu \nu} (x,x') 
 \equiv 
\sum_{\sigma_1=\pm} \sum_{\sigma_2=\pm}
& &\left\{
{\bar \psi}^{(+\omega_1)}_{{\bf k}_1 \sigma_1}(x)
\gamma^{\mu}(c_V-c_A\gamma^5)
\psi^{(-\omega_2)}_{-{\bf k}_2 -\sigma_2}(x) \right.
\nonumber \\
& &
\left. \times
{\bar \psi}^{(-\omega_2)}_{-{\bf k}_2 -\sigma_2}(x')
\gamma^{\nu}(c_V-c_A\gamma^5)
\psi^{(+\omega_1)}_{{\bf k}_1 \sigma_1}(x')
\right\}\;.
\label{Gprel}
\end{eqnarray}
Eq.~(\ref{J}) can be cast in the form
\begin{equation}
{\rm J}^{(p_1 \to p_2)}_{\mu \nu} (x,x')=G_{\rm eff}^2 \;
\frac{u_{\mu}(\tau)u_{\nu} (\tau ')\;
}{u^0(\tau) u^0(\tau ')}
\;e^{i \Delta M (\tau-\tau ')} 
\delta^3({\bf x}-{\bf x}(\tau))\;
\delta^3 ({\bf x}'-{\bf x}(\tau '))
\label{J2}
\end{equation}
by using our current~(\ref{CI}), where  
$\Delta M \equiv M_2-M_1$, while Eq.~(\ref{Gprel}) is 
written as
\begin{eqnarray}
{\rm G}_{{\bf k}_1 {\bf k}_2}^{\mu \nu} (x,x')
=
{\rm tr}
& &\left\{
\gamma^{\mu}(c_V-c_A\gamma^5)
\sum_{\sigma_2=\pm}
\left[ \psi^{(-\omega_2)}_{-{\bf k}_2 -\sigma_2}(x)
{\bar \psi}^{(-\omega_2)}_{-{\bf k}_2 -\sigma_2}(x')
\right] \right.
\nonumber \\
& &\left.
\times \gamma^{\nu}(c_V-c_A\gamma^5)
\sum_{\sigma_1=\pm}
\left[ \psi^{(+\omega_1)}_{{\bf k}_1 \sigma_1}(x')
{\bar \psi}^{(+\omega_1)}_{{\bf k}_1 \sigma_1}(x)
\right]
\right\} \;.
\label{G}
\end{eqnarray}
The summations that appear in Eq.~(\ref{G}) can be calculated
by using modes~(\ref{NM1})-(\ref{NM2}):
\begin{equation}
\sum_{\sigma=\pm} 
\psi^{(\pm \omega)}_{\pm {\bf k} \sigma}(x)
{\bar \psi}^{(\pm \omega)}_{\pm{\bf k} \sigma}(x') =
\frac{(k \hspace{-0.24cm}/ \pm m)}{2 (2\pi)^3 \omega}
 e^{\pm i k^{\lambda} (x-x')_{\lambda}}\; ,
\label{SUM}
\end{equation}
where $k^{\lambda}=(\omega,{\bf k})$ is the emitted fermion's 
four-momentum.
Applying the above expression in Eq.~(\ref{G}), 
and using $\gamma$-matrix trace identities, we obtain
\begin{eqnarray}
{\rm G}_{{\bf k}_1 {\bf k}_2}^{\mu \nu} (x,x')
&=&
\frac{e^{i(k_1+k_2)^{\lambda} (x-x')_{\lambda}}}{4 (2\pi)^6 \omega_1 \omega_2}
\left\{(c_V^2+c_A^2){\rm tr}[\gamma^{\mu} k \hspace{-0.24cm}/_2 \gamma^{\nu}
k \hspace{-0.24cm}/_1]+ 2c_Vc_A {\rm tr}[\gamma^5
\gamma^{\mu} k \hspace{-0.24cm}/_2 \gamma^{\nu}k \hspace{-0.24cm}/_1]
\right.
\nonumber \\
& &\left. -m_1m_2 (c_V^2-c_A^2){\rm tr}[\gamma^{\mu}\gamma^{\nu}]
\right\}
\nonumber \\
&=&
\frac{e^{i(k_1+k_2)^{\lambda} (x-x')_{\lambda}}}{(2\pi)^6 \omega_1 \omega_2}
\left\{
(c_V^2+c_A^2)
\left[2k_1^{(\mu}k_2^{\nu)}-\eta^{\mu\nu} k_1^\alpha {k_2}_\alpha \right]
-m_1m_2(c_V^2-c_A^2)\eta^{\mu\nu} \right.
\nonumber \\
& & \left. +2ic_Vc_A\epsilon^{\mu\nu\alpha\beta} k_{1\alpha} k_{2 \beta} 
\right\}\;,
\label{G4}
\end{eqnarray}
where $\epsilon^{\mu\alpha\nu\beta}$ is the totally skew-symmetric 
Levi-Civita pseudo-tensor (with $\epsilon^{0123}=-1$) and
$
k_1^{(\mu}k_2^{\nu)} \equiv 
(k_1^{\mu}k_2^{\nu}+k_1^{\nu}k_2^{\mu})/2.
$
Letting Eqs.~(\ref{J2}) and (\ref{G4}) into (\ref{dP2}), we  obtain
the differential transition probability
\begin{eqnarray}
\frac{d{\cal P}^{p_1 \to p_2}}{d^3{\bf k}_1 d^3{\bf k}_2}\;&=&\;
\frac{G_{\rm eff}^2}{(2\pi)^6 \omega_1 \omega_2}
\int_{-\infty}^{+\infty} d\tau 
\int_{-\infty}^{+\infty} d\tau'
e^{i\Delta M(\tau-\tau')}\;
e^{i(k_1+k_2)^\lambda [x(\tau)-x(\tau')]_\lambda}
\nonumber \\ 
& &
\times \left\{2
\left[(c_V^2+c_A^2) k_1^{(\mu} k_2^{\nu)} +
ic_Vc_A \epsilon^{\mu\nu\alpha\beta}
k_{1\alpha} k_{2\beta} \right] u_\mu(\tau)u_\nu(\tau') \right.
\nonumber \\
& &\left.
-\left[(c_V^2-c_A^2)m_1m_2 + (c_V^2+c_A^2)k_1^\alpha {k_2}_\alpha \right] 
u^\mu(\tau)u_\mu(\tau')
\right\}\; ,
\label{dP3}
\end{eqnarray}
where we have used that  $d\tau=dt/u^0 $ 

\setcounter{equation}{0}


\section{Uniformly accelerated currents}
\label{accelerated}

The world line of a uniformly accelerated particle with proper
acceleration $a$ can be given in the usual Minkowski
coordinates $(t,{\bf x}) \in {\rm R}^4$ by 
\begin{equation}
x^\mu(\tau)= (a^{-1} \sinh a\tau\;,\;0\;,\;0\;,\;a^{-1} \cosh a\tau) \; .
\label{WL}
\end{equation}
The corresponding four-velocity is 
\begin{equation}
u^\mu(\tau)=(\cosh a\tau\;,\;0\;,\;0\;,\;\sinh a\tau)\;.
\label{UACC}
\end{equation}

Let us now define new coordinates
\begin{equation}
\xi \equiv ({\tau-\tau'})/{2}\;\;\;\;{\rm and}\;\;\;\;
s \equiv ({\tau+\tau'})/{2}\;,
\label{CCG}
\end{equation}
which allows us to rewrite Eq.~(\ref{dP3}) as
\begin{eqnarray}
\frac{d{\cal P}^{p_1 \to p_2}}{d^3{\bf k}_1 d^3{\bf k}_2}\;&=&\;
\frac{2 G_{\rm eff}^2}{(2\pi)^6 \omega_1 \omega_2}
\int_{-\infty}^{+\infty} ds
\int_{-\infty}^{+\infty} d\xi\;
\exp \{2i[\Delta M \xi +  (k_1+k_2)^\lambda u_\lambda(s)\sinh(a\xi)/a] \}
\nonumber \\
& & \times
\left\{
2(c_V^2+c_A^2)k_1^\mu k_2^\nu \left[ u_\mu(s)u_\nu(s) \cosh^2(a\xi) -
a^2  x_\mu(s) x_\nu(s) \sinh^2(a\xi) \right] \right.
\nonumber \\
& & 
- \cosh(2a\xi) \left[
(c_V^2-c_A^2)m_1 m_2 + (c_V^2+c_A^2)(k_1^\alpha {k_2}_\alpha)
\right]
\nonumber \\
& & \left.
+ 2 i a c_Vc_A  \sinh(2a\xi)\;\epsilon_{\mu\nu\alpha\beta}\;x^\mu(s) u^\nu(s)
k_1^\alpha k_2^\beta
\right\}
\;,
\label{dP4}
\end{eqnarray}
where we have used
$
[x(\tau)-x(\tau')]^\mu=2a^{-1}\sinh(a\xi) u^\mu (s),
$
$
u^\mu(\tau)=\cosh(a\xi)u^\mu(s)+a\sinh(a\xi)x^\mu(s),
$
$
u^\mu(\tau')=\cosh(a\xi)u^\mu(s)-a\sinh(a\xi)x^\mu(s),
$
and
$
u^\mu(\tau)u_\mu(\tau')= \cosh(2a\xi).
$

In order to decouple the integrals in Eq.~(\ref{dP4}), 
let us make the following change in the momentum variable:
\begin{equation}
k^\mu \to {\tilde k}^\mu = ( {\tilde \omega}, {\tilde {\bf k}})
\equiv (\;k^\lambda u_\lambda(s)\;,\;
 k^x\;,\;k^y\;,\;-ak^\lambda
x_\lambda(s)\;)\;.
\label{CMC}
\end{equation}
Using Eqs.~(\ref{WL}) and (\ref{UACC}) we can verify explicitly that the
transformation (\ref{CMC}) corresponds to a boost in the 
{\it z}-direction. Indeed,
${\tilde k}^\mu$ are the components of the emitted fermion's 
four-momentum in the inertial frame instantaneously at rest with the
current at the proper time $s$. Hence the transition probability per 
proper time $\Gamma^{p_1 \to p_2} \equiv {d{\cal P}^{p_1 \to p_2}}/{ds}$
for process (\ref{p1p2}) can be written from Eq.~(\ref{dP4}) as
\begin{eqnarray}
\frac{d\Gamma^{p_1 \to p_2}}{d^3{\tilde {\bf k}}_1
d^3{\tilde {\bf k}}_2}\;&=&\;
\frac{2 G_{\rm eff}^2}{(2\pi)^6 {\tilde \omega}_1 {\tilde \omega}_2}
\int_{-\infty}^{+\infty} d\xi\;
\exp \{ 2i[\Delta M \xi + a^{-1} \sinh (a\xi) ({\tilde \omega}_1+
{\tilde \omega}_2)]\}
\nonumber \\
& &
\times
\left\{
(c_V^2+c_A^2)({\tilde \omega}_1{\tilde \omega}_2+
{\tilde k}^z_1{\tilde k}^z_2)
-2ic_Vc_A \sinh(2a\xi)({\tilde {\bf k}}_1 \times {\tilde {\bf k}}_2)^z
\right.
\nonumber \\
& &\left.
+ \left[
(c_V^2+c_A^2)({\tilde {\bf k}}^\bot_1 \cdot {\tilde {\bf k}}^\bot_2)
-(c_V^2-c_A^2)m_1 m_2\right]\cosh(2a\xi) \right\}\; ,
\label{dG1}
\end{eqnarray}
where 
${\tilde {\bf k}}_1 \times {\tilde {\bf k}}_2$ is the
usual three-vector product and ${\tilde {\bf k}}^\bot_1 
\cdot {\tilde {\bf k}}^\bot_2
\equiv {\tilde k}^x_1{\tilde k}^x_2+
{\tilde k}^y_1{\tilde k}^y_2$.
In order to integrate Eq.~(\ref{dG1}), it is convenient to 
use spherical coordinates in the momenta space 
$
({\tilde k}\in {\rm R}^+,
 {\tilde \theta}\in [0,\pi],
 {\tilde \phi}\in [0,2\pi))
$, 
where
$
{\tilde k}^x = {\tilde k} \sin {\tilde \theta} \cos {\tilde \phi}
$,
$
{\tilde k}^y = {\tilde k} \sin {\tilde \theta} \sin {\tilde \phi}
$,
$
{\tilde k}^z = {\tilde k} \cos {\tilde \theta}
$,
and the following change of integration variable: 
$\xi \to \lambda \equiv e^{a\xi}$.
By using expression (3.471.10) of Ref.~\cite{GRAD}, we obtain
\begin{eqnarray}
\frac{d\Gamma^{p_1 \to p_2}}{d^3{\tilde {\bf k}}_1
d^3{\tilde {\bf k}}_2}\;&=&\;
\frac{4 G_{\rm eff}^2 \;e^{-\pi \Delta M/a}}
{(2\pi)^6 {\tilde \omega}_1 {\tilde \omega}_2a}
\left\{
(c_V^2+c_A^2)({\tilde \omega}_1{\tilde \omega}_2+
{\tilde k}_1{\tilde k}_2 \cos {\tilde \theta}_1 
\cos {\tilde \theta}_2) \; K_{2i \Delta M/a} 
\left[
2({\tilde \omega}_1+{\tilde \omega}_2)/a 
\right] \right.
\nonumber \\
& &
+2c_Vc_A{\tilde k}_1{\tilde k}_2 \sin {\tilde \theta}_1 \sin {\tilde \theta}_2
\sin ({\tilde \phi}_1-{\tilde \phi}_2)\; {\rm Im} 
\left\{ K_{2+2i \Delta M/a} 
\left[
2({\tilde \omega}_1+{\tilde \omega}_2)/a 
\right] \right\}
\nonumber \\
& &
+ \left[
(c_V^2-c_A^2) m_1 m_2- 
(c_V^2+c_A^2){\tilde k}_1{\tilde k}_2\sin {\tilde \theta}_1 
\sin {\tilde \theta}_2 \cos ({\tilde \phi}_1-{\tilde \phi}_2)
\right] 
\nonumber \\
& & \left. \times
{\rm Re}
\left\{ K_{2+2i \Delta M/a} 
\left[
2({\tilde \omega}_1+{\tilde \omega}_2)/a 
\right] \right\}
\right\}\;,
\label{dG3}
\end{eqnarray}
where ${\rm Re} \{z\}$ and ${\rm Im}\{z\}$ are the real and imaginary parts 
of a complex number $z$, respectively, and $K_{\nu} (z)$ is the modified
Bessel function.

We note that the uncorrelated emission of $f_1$ and ${\bar f}_2$
is {\it spherically symmetric} in the instantaneously 
comoving frame. This can be seen by tracing out (i.e., integrating)
one of the momentum variables in Eq.~(\ref{dG3}): 
\begin{eqnarray}
\frac{d\Gamma^{p_1 \to p_2}}{d^3{\tilde {\bf k}}_j} \;&=&\;
\frac{8 G_{\rm eff}^2 \;e^{-\pi \Delta M/a}}
{(2\pi)^5 {\tilde \omega}_j a}
\int_{0}^\infty d{\tilde k}_l\; \frac{{\tilde k}_l^2}{{\tilde \omega}_l}
\left\{
(c_V^2+c_A^2){\tilde \omega}_1{\tilde \omega}_2
\;K_{2i \Delta M/a} 
\left[
2({\tilde \omega}_1+{\tilde \omega}_2)/a 
\right] \right.
\nonumber \\
& & \left. +
(c_V^2-c_A^2) m_1 m_2
\;{\rm Re}
\left\{ K_{2+2i \Delta M/a} 
\left[
2({\tilde \omega}_1+{\tilde \omega}_2)/a 
\right] \right\}
\right\}\;,
\label{dG4}
\end{eqnarray}
and noting that this expression is  independent of 
$({\tilde \theta}_j,{\tilde \phi}_j)$, where $j,l=1$ and $2$ 
are associated with particles $f_1$ and
${\bar f}_2$. The energy distribution of emitted particles is given by:
\begin{eqnarray}
\frac{d\Gamma^{p_1 \to p_2}}{d{\tilde \omega}_j}\;&=&\;
\frac{G_{\rm eff}^2 \;e^{-\pi \Delta M/a}}
{\pi^4 a}
\sqrt{{\tilde \omega}_j^2-m_j^2}
\int_{m_l}^\infty d{\tilde \omega}_l\;
\sqrt{{\tilde \omega}_l^2-m_l^2}\;
\nonumber \\
& & 
\times
\left\{
(c_V^2+c_A^2){\tilde \omega}_1{\tilde \omega}_2
\;K_{2i \Delta M/a} 
\left[
2({\tilde \omega}_1+{\tilde \omega}_2)/a 
\right] \right.
\nonumber \\
& & \left.
+(c_V^2-c_A^2) m_1 m_2
\;{\rm Re}
\left\{ K_{2+2i \Delta M/a} 
\left[
2({\tilde \omega}_1+{\tilde \omega}_2)/a 
\right] \right\}
\right\}\; .
\label{dG5}
\end{eqnarray}
The total transition rate is given by
\begin{eqnarray}
\Gamma^{p_1 \to p_2} \;&=&\;
\frac{G_{\rm eff}^2 \;e^{-\pi \Delta M/a}}
{\pi^4 a}
\int_{m_1}^\infty d{\tilde \omega}_1
\int_{m_2}^\infty d{\tilde \omega}_2\;
\sqrt{{\tilde \omega}_1^2-m_1^2}
\sqrt{{\tilde \omega}_2^2-m_2^2}
\nonumber \\
& & \times
\left\{
(c_V^2+c_A^2){\tilde \omega}_1{\tilde \omega}_2\;
K_{2i \Delta M/a} 
\left[
2({\tilde \omega}_1+{\tilde \omega}_2)/a 
\right] \right.
\nonumber \\
& &  \left.
+
(c_V^2-c_A^2) m_1 m_2
\;{\rm Re}
\left\{ K_{2+2i \Delta M/a} 
\left[
2({\tilde \omega}_1+{\tilde \omega}_2)/a 
\right] \right\}
\right\}
\label{GN}
\end{eqnarray}
while the emitted power can be estimated by
\begin{eqnarray}
{\cal W}_j^{p_1 \to p_2} \;&=&\;
\frac{G_{\rm eff}^2 \;e^{-\pi \Delta M/a}}
{\pi^4 a}
\int_{m_1}^\infty d{\tilde \omega}_1
\int_{m_2}^\infty d{\tilde \omega}_2\;
{\tilde \omega}_j
\sqrt{{\tilde \omega}_1^2-m_1^2}
\sqrt{{\tilde \omega}_2^2-m_2^2}
\nonumber \\
& & \times
\left\{
(c_V^2+c_A^2){\tilde \omega}_1{\tilde \omega}_2\;
K_{2i \Delta M/a} 
\left[
2({\tilde \omega}_1+{\tilde \omega}_2)/a 
\right] \right.
\nonumber \\
& & \left.
+
(c_V^2-c_A^2) m_1 m_2
\;{\rm Re}
\left\{ K_{2+2i \Delta M/a} 
\left[
2({\tilde \omega}_1+{\tilde \omega}_2)/a 
\right] \right\}
\right\}\; .
\label{WN}
\end{eqnarray}

Assuming that $f_1$ {\it or}  ${\bar f}_2$ is a massless particle, 
we can perform explicitly the integrals that appear in Eqs.~(\ref{GN}) 
and (\ref{WN}). For this purpose, we make the change of variables 
$({\tilde \omega}_1,{\tilde \omega}_2) \to (\rho, \zeta)$, where
\begin{equation}
\rho \equiv {{\tilde \omega}_l}/{{\tilde \omega}_i}+1 
\;\;{\rm and}
\;\;
\zeta \equiv {{\tilde \omega}_i^2}/{m^2}\; ,
\label{CV2}
\end{equation}
and here we label the massless and massive (with mass $m$) 
particles with  $l$ and $i$ indices, respectively. 
Applying (\ref{CV2}) in Eqs.~(\ref{GN}) and (\ref{WN})
with $m_l=0$, we have
\begin{eqnarray}
\Gamma^{p_1 \to p_2} \;&=&\;
\frac{G_{\rm eff}^2 (c_V^2+c_A^2) m^6 }
{2\pi^4 a e^{\pi \Delta M/a}}
\int_{1}^\infty d\rho \;
(\rho-1)^2
\int_{1}^\infty d\zeta \;
\zeta^{3/2}(\zeta-1)^{1/2}
\nonumber \\
& &
\times K_{2i\Delta M/a} \left[ 2m\rho \;\zeta^{1/2}/a \right]\;,
\label{GA1}
\end{eqnarray}
\begin{eqnarray}
{\cal W}_{\rm massive}^{p_1 \to p_2}  \;&=&\;
\frac{G_{\rm eff}^2 (c_V^2+c_A^2) m^7 }
{2\pi^4 a e^{\pi \Delta M/a}}
\int_{1}^\infty d\rho \;
(\rho-1)^2
\int_{1}^\infty d\zeta \;
\zeta^2(\zeta-1)^{1/2}
\nonumber \\
& &
\times K_{2i\Delta M/a} \left[ 2m\rho \;\zeta^{1/2}/a \right]\;,
\label{WJ1}
\end{eqnarray}
and
\begin{eqnarray}
{\cal W}_{\rm massless}^{p_1 \to p_2}  \;&=&\;
\frac{G_{\rm eff}^2 (c_V^2+c_A^2) m^7 }
{2\pi^4 a e^{\pi \Delta M/a}}
\int_{1}^\infty d\rho \;
(\rho-1)^3
\int_{1}^\infty d\zeta \;
\zeta^2(\zeta-1)^{1/2}
\nonumber \\
& &
\times K_{2i\Delta M/a} \left[ 2m\rho \;\zeta^{1/2}/a \right]\;.
\label{WL1}
\end{eqnarray}
By using Eq. (6.592.4) of Ref.~\cite{GRAD} to perform 
the $\zeta$-integration
in Eqs.~(\ref{GA1})-(\ref{WL1}), we obtain
\begin{eqnarray}
\Gamma^{p_1 \to p_2} \;&=&\;
\frac{G_{\rm eff}^2 (c_V^2+c_A^2)  m^3 a^2}
     {8 \pi^{7/2} e^{\pi \Delta M/a}}\;
\int_{1}^\infty d\rho \;
\left( \rho^{-1} -2\rho^{-2} + \rho^{-3} \right)
\nonumber \\
& &
\times 
G_{1\;3}^{3\;0} \left( \frac{m^2 \rho^2}{a^2} \left|
\begin{array}{l}
\;\;\;0\\  
-{3}/{2}\;,\; {3}/{2}+i {\Delta M}/{a}\;,\;
{3}/{2}-i{\Delta M}/{a}
\end{array}
\right.
\right)
\;,
\label{GA2}
\end{eqnarray}
\begin{eqnarray}
{\cal W}_{\rm massive}^{p_1 \to p_2} \;&=&\;
\frac{G_{\rm eff}^2 (c_V^2+c_A^2) m^3 a^3}{8 \pi^{7/2} \;
e^{\pi \Delta M/a}}\;
\int_{1}^\infty d\rho \;
\left( \rho^{-2} -2\rho^{-3} + \rho^{-4} \right)
\nonumber \\
& &
\times 
G_{1\;3}^{3\;0} \left( \frac{m^2 \rho^2}{a^2} \left|
\begin{array}{l}
\; \; \;0\\ 
-{3}/{2}\;,\;2+i{\Delta M}/{a}\;,\;
2-i{\Delta M}/{a}
\end{array}
\right.
\right)
\;,
\label{WJ2}
\end{eqnarray}
\begin{eqnarray}
{\cal W}_{\rm massless}^{p_1 \to p_2} \;&=&\;
\frac{G_{\rm eff}^2 (c_V^2+c_A^2) m^3 a^3}{8 \pi^{7/2} \;e^{\pi 
\Delta M/a}}\;
\int_{1}^\infty d\rho \;
\left( \rho^{-1} -3\rho^{-2} + 3\rho^{-3} -\rho^{-4}\right)
\nonumber \\
& &
\times 
G_{1\;3}^{3\;0} 
\left(  \frac{m^2 \rho^2}{a^2} \left|
\begin{array}{l}
\;\;\;0\\ 
-{3}/{2}\;,\;2+i{\Delta M}/{a}\;,\;2-i {\Delta M}/{a}
\end{array}
\right.
\right)
\;,
\label{WL2}
\end{eqnarray}
where $ G_{p\;q}^{m\;n} \left( x \vert 
^{a_1,...,a_p}_{b_1,...,b_q}
\right)$ are the Meijer's $G$-functions (see  Ref.~\cite{GRAD}
for their definition and properties). Defining $v\equiv \rho^2$ in 
Eqs.~(\ref{GA2})-(\ref{WL2}), and using Eq.~(7.811.3) of 
Ref.~\cite{GRAD}, we can integrate these expressions. The Meijer's
$G$-function sums that appear as a result can be simplified by using
their properties. Eventually, we obtain   
\begin{equation}
\Gamma^{p_1 \to p_2} =
\frac{G_{\rm eff}^2 (c_V^2+c_A^2) m^3 a^2 }{32 \pi^{7/2} 
\;e^{\pi \Delta M/a}}\;
G^{4\;0}_{2\;4}
\left(
\frac{m^2}{a^2}
\left|
\begin{array}{l}
{3}/{2}\;,\;\;\;\;\;2\\  
{1}/{2}\;,\;-{3}/{2}\;,\;{3}/{2}+i{\Delta M}/{a}\;,\;
{3}/{2}-i{\Delta M}/{a}
\end{array}
\right.
\right),
\label{GA4}
\end{equation}
\begin{equation}
{\cal W}_{\rm massive}^{p_1 \to p_2} =
\frac{G_{\rm eff}^2 (c_V^2+c_A^2) m^3 a^3}{32 \pi^{7/2} 
\;e^{\pi \Delta M/a}}\;
G^{5\;0}_{3\;5}
\left(
\frac{m^2}{a^2}
\left|
\begin{array}{l}
\;\;0\;\;,\;2\;,\;\;\;\;{5}/{2}\\  
{1}/{2}\;,\;1\;,\;-{3}/{2}\;,\;2+i{\Delta M}/{a}\;,\;
2-i{\Delta M}/{a}
\end{array}
\right.
\right),
\label{WJ4}
\end{equation}
\begin{equation}
{\cal W}_{\rm massless}^{p_1 \to p_2}=
\frac{3\;G_{\rm eff}^2 (c_V^2+c_A^2)m^3 a^3}{64 \pi^{7/2} 
\;e^{\pi \Delta M/a}}\;
 G^{4\;0}_{2\;4}
\left(
\frac{m^2}{a^2}
\left|
\begin{array}{l}
\;\;2\;\;,\;\;\;\;{5}/{2}\\  
{1}/{2}\;,\;-{3}/{2}\;,\;2+i{\Delta M}/{a}\;,\;
2-i{\Delta M}/{a}
\end{array}
\right.
\right)\;.
\label{WL4}
\end{equation}
\begin{figure}
\begin{center}
\mbox{\epsfig{file=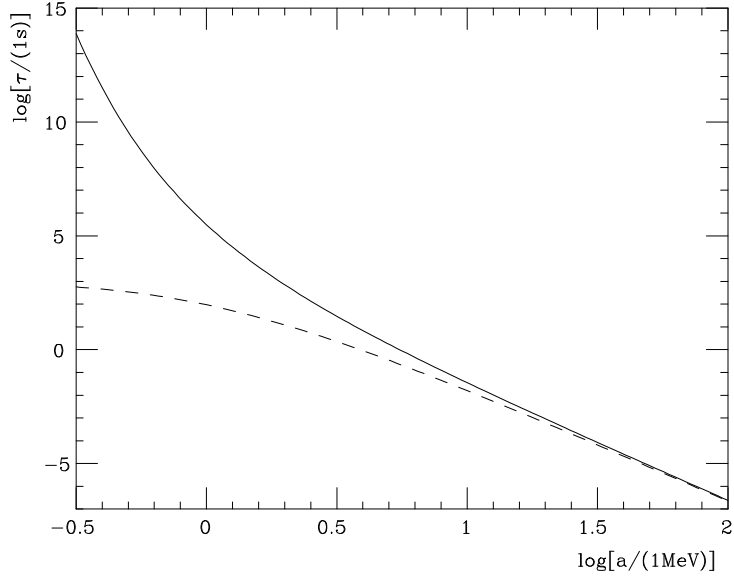,width=0.6\textwidth,angle=90}}
\end{center}
\vskip -1 cm
\caption{The mean proper lifetime of protons, $\tau_p$ (full line),
and neutrons, $\tau_n$ (dashed line), are plotted 
as functions of their proper acceleration $a$. Note that
$\tau_p \to + \infty$ and 
$\tau_n \to 887$ s as $a \to 0$.
For accelerations 
$a  \gg a_c \equiv 2\pi \Delta M\approx 8$ MeV 
we have that $\tau_p \approx \tau_n $. }
\label{fig1}
\end{figure}

In the case where both $f_1$ and ${\bar f}_2$ are massless particles, 
this is more convenient to obtain the total transition rate by 
first integrating in momenta ${\tilde {\bf k}}_1$ and ${\tilde {\bf k}}_2$.
Thus, we first write [see Eq.~(\ref{dG1})]
\begin{equation}
\Gamma^{p_1 \to p_2}\;=\;
\frac{G_{\rm eff}^2 (c_V^2+c_A^2)}{2\pi^4}
\int_{-\infty}^{+\infty}d\xi\;
e^{2i\Delta M \xi}
\left\{
\int_0^\infty d{\tilde \omega}\;
{\tilde \omega}^2
\exp \left[
\frac{2i{\tilde \omega}}{a}\left(\sinh a\xi +i\epsilon\right)\right]
\right\}^2\;,
\label{GAIII1}
\end{equation}
where $\epsilon >0$ is a regulator that ensures the convergence of the 
frequency integral above. The corresponding total emitted power is
\begin{eqnarray}
{\cal W}^{p_1 \to p_2}\;&=&\;
\frac{G_{\rm eff}^2 (c_V^2+c_A^2)}{\pi^4}
\int_{-\infty}^{+\infty}d\xi\;
e^{2i\Delta M \xi}
\int_0^\infty d{\tilde \omega}_1\;
{\tilde \omega}_1^3
\exp \left[
\frac{2i{\tilde \omega}_1}{a}\left(\sinh a\xi +i\epsilon\right)\right]
\nonumber \\
& &\times
\int_0^\infty d{\tilde \omega}_2\;
{\tilde \omega}_2^2
\exp \left[
\frac{2i{\tilde \omega}_2}{a}\left(\sinh a\xi +i\epsilon\right)\right]
\;.
\label{WIII1}
\end{eqnarray}
By performing the frequency integrals and defining the new variable
$w\equiv e^{a\xi}$, Eqs.~(\ref{GAIII1})-(\ref{WIII1}) become
\begin{equation}
\Gamma^{p_1 \to p_2}\;=\;
-\;\frac{2G_{\rm eff}^2 (c_V^2+c_A^2)\;a^5}{\pi^4}
\int_{0}^{\infty}dw\;
\frac{w^{5+2i\Delta M/a}}
{(w^2-1+2i\epsilon w)^6}\;,
\label{GAIII3}
\end{equation}
\begin{equation}
{\cal W}^{p_1 \to p_2}\;=\;
-\;\frac{12iG_{\rm eff}^2 (c_V^2+c_A^2)\;a^6}{\pi^4}
\int_{0}^{\infty}dw\;
\frac{w^{6+2i\Delta M/a}}
{(w^2-1+2i\epsilon w)^7}\;.
\label{WIII3}
\end{equation}
\begin{figure}
\begin{center}
\mbox{\epsfig{file=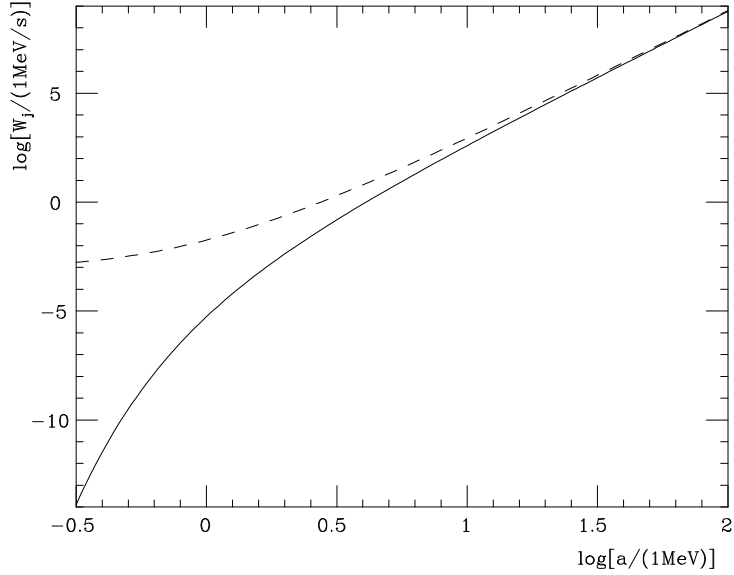,width=0.6\textwidth,angle=90}}
\end{center}
\vskip -1 cm
\caption{${\cal W}_{j}^{p\to n}$ and  ${\cal W}_{j}^{n\to p}$ 
are plotted in full and dashed lines, respectively, for $j=e ,\nu$,
as  functions of the nucleon proper accelerations.
Our numerical results suggest that 
${\cal W}_{e}^{p\to n} = {\cal W}_{\nu}^{p\to n}$
and 
${\cal W}_{e}^{n\to p} = {\cal W}_{\nu}^{n\to p}$.}
\label{fig2}
\end{figure}
Solving the integrals that appear in Eqs.~(\ref{GAIII3}) and (\ref{WIII3})
(see Appendix), we obtain
\begin{equation}
\Gamma^{p_1 \to p_2}\;=\;
\frac{G_{\rm eff}^2 (c_V^2+c_A^2)}{60\;\pi^3}
\left(\frac{4\;a^4 \Delta M + 5\;a^2 \Delta M^3 +\Delta M^5}
{e^{2\pi\Delta M/a}-1}\right)\;
\label{GAIII4}
\end{equation}
and
\begin{equation}
{\cal W}^{p_1 \to p_2}\;=\;
\frac{G_{\rm eff}^2 (c_V^2+c_A^2)}{3840\;\pi^3}
\left(
\frac{225\;a^6 +1036 \;a^4 \Delta M ^2 + 560\;a^2\Delta M ^4+
64\; \Delta M^6}{e^{2\pi \Delta M/a}+1}
\right) \;.
\label{WIII4}
\end{equation}
In the next sections we use these formulas to 
investigate some selected reactions.

\setcounter{equation}{0}


\section{Accelerated proton and neutron decay}
\label{decay}

Let us now consider the processes
\begin{equation}
p \to n \;e^+  \nu_e\;
\label{pn}
\end{equation}
and
\begin{equation}
n \to p \;e^- \bar \nu_e\;
\label{np}
\end{equation}
for uniformly accelerated protons and neutrons, respectively.
We will assume the neutrino mass to vanish because even if this
is not so, it would be neglectable in comparison 
with any other energy scale involved in the problem.  The effective coupling 
constant $G_{\rm eff}= G_{pn}$ for processes (\ref{pn})-(\ref{np})
is obtained by imposing that the 
mean proper lifetime of inertial neutrons is $887$ s~\cite{PDG}, i.e.,
\begin{equation}
\Gamma^{n\to p}_{in} \equiv
\Gamma^{n\to p}(a\to 0)=1/887\;{\rm s}^{-1}\;.
\label{GEFF}
\end{equation}

This phenomenological procedure has the advantage of by passing any 
uncertainties on the influence of the nucleon inner structure.
For sake of convenience, we take the $a\to 0$ limit in Eq.~(\ref{dG1})
rather than in Eq.~(\ref{GA4}), obtaining
\begin{eqnarray}
\frac{d\Gamma^{n \to p}_{in}}{d^3{\tilde {\bf k}}_e
d^3{\tilde {\bf k}}_\nu}\;&=&\;
\frac{4 G_{pn}^2}{(2\pi)^6 {\tilde \omega}_e {\tilde \omega}_\nu}
\int_{-\infty}^{+\infty} d\xi\;
e^{2i\xi(\Delta M  + {\tilde \omega}_e+
{\tilde \omega}_\nu)}\;
({\tilde \omega}_e{\tilde \omega}_\nu+
{\tilde {\bf k}}_e \cdot {\tilde {\bf k}}_\nu)\;
\nonumber \\
&=&\; 
\frac{2 G_{pn}^2}{(2\pi)^5 }
\; \left( 1+
\frac{{\tilde {\bf k}}_e \cdot {\tilde {\bf k}}_\nu}
{{\tilde \omega}_e {\tilde \omega}_\nu}
\right) \;
\delta ({\tilde \omega}_e+
{\tilde \omega}_\nu-\Delta M )\;,
\label{dGIN}
\end{eqnarray}
where we have used $c_V=c_A=1$ \cite{CQ} since only left-handed 
massless neutrinos
are known to exist.
After integrating Eq.~(\ref{dGIN}) in angular coordinates and in
$\tilde \omega_e$, we find
\begin{equation}
\Gamma^{n\to p}_{in} =
\frac{G_{pn}^2}{\pi^3}
\int_0^{\Delta M-m_e} d{\tilde \omega}_\nu\;{\tilde \omega}_\nu^2
\left(
\Delta M -{\tilde \omega}_\nu
\right)
\sqrt{
\left(
\Delta M -{\tilde \omega}_\nu
\right) ^2-m_e^2}\;.
\label{dGIN2}
\end{equation}

Evaluating numerically Eq.~(\ref{dGIN2}) with 
$m_e= 0.511 \; {\rm MeV}$, and
$\Delta M= (m_n-m_p)= 1.29 \;{\rm MeV}$, we end up with
$
\Gamma^{n\to p}_{in}\;=\;
1.81 \times 10^{-3} \;G_{pn}^2 \; {\rm MeV}^5
$.
Hence by imposing condition (\ref{GEFF}), we obtain 
$
G_{pn}\;=\;1.74\;G_F\;,
$
where $G_F \equiv 1.166\times 10^{-5}\;{\rm GeV}^{-2}$
is the Fermi coupling constant~\cite{PDG}.
Now we are able to use Eq.~(\ref{GA4}) to plot in
Fig.~\ref{fig1} the proton and neutron mean proper lifetimes $\tau_p (a)= 
\left( \Gamma^{p\to n} \right) ^{-1}$ and $\tau_n (a)= 
\left( \Gamma^{n\to p} \right) ^{-1}$, respectively. Let us note that
\begin{equation}
\tau_n (a)= e^{-2\pi |\Delta M|/a}\;\tau_p (a)\;.
\label{TPTN}
\end{equation}
We have only considered accelerations $a \ll m_p=938$ MeV in order to
respect our no-recoil condition (see Sec.~\ref{DNC}). We call attention
to the fact that for accelerations 
$a \gg a_c \equiv 2 \pi |\Delta M| \approx 8$ MeV,
we have $\tau_p (a) \approx \tau_n(a)$. 
This is easier to understand in the co-accelerated frame with the current,
where (according to the FDU effect~\cite{FD,U}) a thermal bath of Rindler 
particles 
with temperature $T_{FDU} = a/2\pi$ is ``attached'' to the current. Thus,
for $a \gg a_c$ we have $T_{FDU} \gg |\Delta M|$, 
which leads both nucleons to behave similarly. 
(See Ref.~\cite{MV} for a more comprehensive discussion on this issue.)

In order to estimate how much energy is 
carried out in form of leptons, we may use Eqs.~(\ref{WJ4}) and (\ref{WL4})
to obtain ${\cal W}_{j}^{p\to n} $ and 
${\cal W}_{j}^{n\to p} = e^{2 \pi |\Delta M|/a}\; {\cal W}_{j}^{p\to n} $ 
for $j= e, \nu$.
Although ${\cal W}_{e}^{p\to n} $ and $ {\cal W}_{\nu}^{p\to n}$ 
(as well as ${\cal W}_{e}^{n\to p} $ and $ {\cal W}_{\nu}^{n\to p}$)
are not manifestly identical, they seem to be according to 
Fig.~\ref{fig2}.

\begin{figure}
\begin{center}
\mbox{\epsfig{file=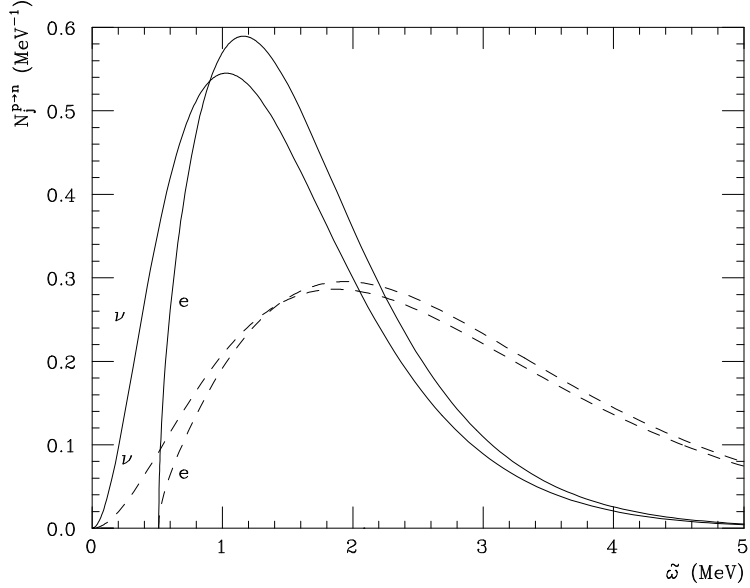,width=0.6\textwidth,angle=90}}
\end{center}
\vskip -1 cm
\caption{The normalized energy distribution of emitted positrons,
${\cal N}_e^{p \to n}$, and neutrinos, 
${\cal N}_{\nu}^{p \to n}$, are plotted for two values of the
proton's proper acceleration: $a=1.0$ MeV (full line) and 
$2.0$ MeV (dashed line). Note that the typical energy
of the emitted particles (in the inertial frame instantaneously
at rest with the proton) is given by ${\tilde \omega} \approx a$.}
\label{fig3}
\end{figure}

In order to investigate the energy distribution of the emitted
leptons, let us define the normalized energy distribution
\begin{equation} 
{\cal N}^{p_1 \to p_2}_j\equiv
\frac{1}{\Gamma^{p_1\to p_2}}
\frac{d\Gamma^{p_1\to p_2}}{d{\tilde \omega}_j}\;
\label{NED}
\end{equation} 
with $j=e, \nu$, where ${d\Gamma^{p_1\to p_2}}/{d{\tilde \omega}_{j}}$ is
defined in Eq.~(\ref{dG5}). Note that 
$ 
{\cal N}^{p \to n}_j={\cal N}^{n \to p}_j\; .
$
In Fig.~\ref{fig3} we plot the distributions ${\cal N}^{p \to n}_j$
for two values of acceleration: $a=1.0\;{\rm MeV}\;{\rm and}
\;2.0\;{\rm MeV}$. We see that the typical energy (in the inertial
frame instantaneously at rest with the nucleon) of the
emitted electrons and neutrinos is ${\tilde \omega}\approx a$, 
which justifies our no-recoil condition.

In order to roughly estimate how small is the proper lifetime of circularly 
moving protons at LHC/CERN we use directly Eq.~(\ref{TPTN})  with 
$a=a_{_{\rm LHC}}\approx 10^{-8}\;{\rm MeV}$
for the proton's proper acceleration, obtaining
$
\tau_p(a_{_{\rm LHC}})\approx 10^{3\times 10^8}{\rm yr} (!)\;,
$
where we have used that $\tau_n\left( a\ll m_e,|\Delta M| \right)\approx
10^3$ s. Although Eq.~(\ref{TPTN}) was derived assuming uniformly accelerated 
motion, this should not be seen as a major problem:  
Because of the huge proper lifetime  
obtained for the proton, our estimation turns out to be non-sensitive 
up to an inaccuracy of hundreds of thousands of orders 
of magnitude (which should not be the case). 

Astrophysics seems to provide much more suitable conditions for the 
observation of the decay of accelerated protons. Although our decay rate 
(\ref{GA4}) was obtained considering uniformly accelerated protons, let
us assume that this is approximately valid for circularly moving protons
with proper acceleration $a \gg \Delta M, 1/R$, where $R$ is
the local curvature radius of the proton trajectory. Indeed we can test
this assumption, e.g., for two-level scalar systems, whose excitation rates,
at the tree level, are given by~\cite{TAK}
\begin{equation}
\Gamma_{\rm lin}\;=\;
\frac{c_0^2}{2\pi}\frac{\Delta E}{e^{2\pi \Delta E /a}-1}
\label{GLIN}
\end{equation}
and
\begin{equation}
\Gamma_{\rm cir}\;=\;
\frac{c_0^2}{2\pi}\frac{a\;e^{-\sqrt{12}\Delta E /a}}{2\; \sqrt{12}}
\label{GCIR}
\end{equation}
for uniformly accelerated and circularly moving relativistic sources, 
respectively, where $c_0$ is a small coupling constant 
and $\Delta E$ is the two-level system energy gap. Note that 
in the limit $a\gg \Delta E$, Eqs.~(\ref{GLIN}) and (\ref{GCIR}) give us
$\Gamma_{\rm lin}/\Gamma_{\rm cir}=1.103$. 
In order to illustrate an astrophysical situation
where process~(\ref{pn}) may be of some importance, 
let us consider a cosmic ray proton
with energy $E_p = \gamma m_p \approx 1.6 \times 10^{14}$ eV 
under the influence of a magnetic field $B \approx 10^{14}$ Gauss
of a typical pulsar.  Protons under these
conditions have proper accelerations of $a_B = \gamma
e B/m_p \approx 110$ MeV $\gg |\Delta M|$. 
For practical purposes the acceleration of the proton will be 
assumed as constant along the process. 
For the chosen values of $E_p$ and $B$, the proton is confined
in a cylinder with typical radius 
$R \approx \gamma^2/a_B \approx 5\,\cdot \, 10^{-3} \; {\rm cm}
\ll l_B , $ where $l_B$ is the typical
size of the magnetic field region.
 According to  Eq.~(\ref{GA4}) we obtain $\tau_p \approx 10^{-7}$ s.
As a result, protons would have a ``laboratory'' mean lifetime of
$t_p = \gamma \tau_p \approx 10^{-1}$ s. For $l_B \approx 10^7$ cm,
we  obtain that less than
$| \Delta N_p /N_p | =(1-e^{-l_B/t_p}) \approx l_B/t_p \approx 1\%$
of the protons would decay via (\ref{pn}). We note that we did not take
into account the influence of the magnetic field on the emitted
positron. Clearly a more precise estimation should take into account this
effect as well as other ones as, e.g., the non-uniformity of the magnetic 
field and energy losses through electromagnetic sinchrotron radiation. 
The last one in particular may not be a problem since energy may be furnished
to the proton from dynamo processes.
A more careful analysis of such astrophysical issues would be welcome but
this is beyond the scope of the present field-theoretical investigation.

\setcounter{equation}{0}


\section{Neutrino emission from uniformly accelerated electrons}
\label{Neutrino emission}

In this section, we will consider the emission of
neutrinos from accelerated electrons:
\begin{equation}
e^- \to e^-\; \nu_e \bar \nu_e\;.
\label{ee}
\label{equation}
\end{equation}
The  description of the creation of neutrino-antineutrino pairs by
electrons in an external electromagnetic field in the context of
the standard model is contained in Sec.~6.1 of Ref.~\cite{BKS}.
Here we analyze this process for uniformly accelerated electrons 
by using the formulas derived in 
Sec.~IV where both emitted fermions are massless.  
From Eqs.~(\ref{GAIII4}) and
(\ref{WIII4}) we get for the emission rate of $ \nu_e {\bar \nu}_e $ pairs 
\begin{equation}
\Gamma_{\nu {\bar \nu}}=
\frac{G_{e\nu}^2\;a^5}{15 \pi^4}\;,
\label{GNU}
\end{equation}
and for the  {\it total} radiated power 
\begin{equation}
{\cal W}_{\nu {\bar \nu}}=
\frac{15\;G_{e\nu}^2\;a^6}{256 \pi^3}\;,
\label{WNU}
\end{equation}
where we have used $\Delta M =0$, $c_V=c_A=1$ and 
$G_{e\nu}$ is the corresponding effective coupling constant.
\begin{figure}
\begin{center}
\mbox{\epsfig{file=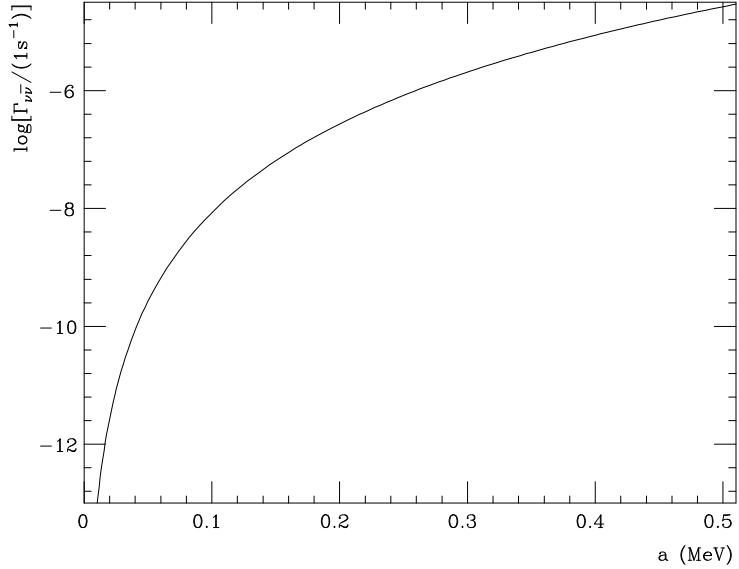,width=0.6\textwidth,angle=90}}
\end{center}
\vskip -1 cm
\caption{The emission probability per proper time of $\nu_e {\bar \nu}_e$ 
pairs is plotted  for $a\leq m_e$.}
\label{fig4}
\end{figure}

\begin{figure}
\begin{center}
\mbox{\epsfig{file=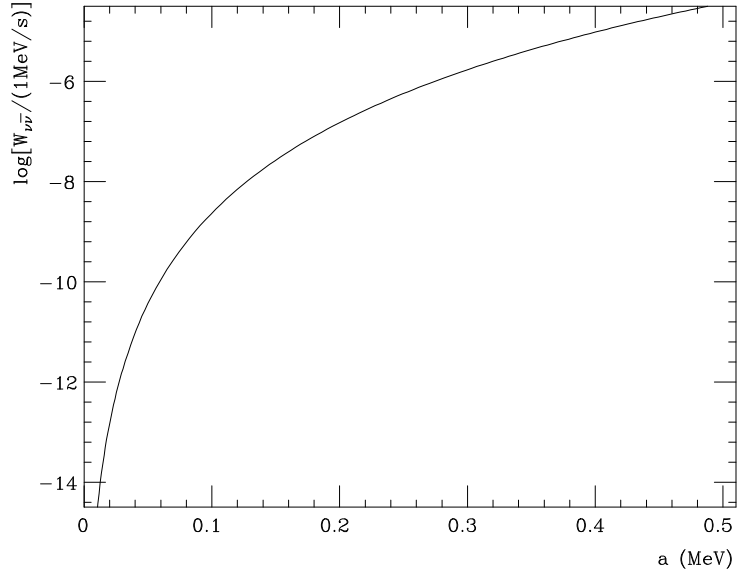,width=0.6\textwidth,angle=90}}
\end{center}
\vskip -1 cm
\caption{ The total radiated power in form of neutrinos is plotted for
$a \leq m_e$.}
\label{fig5}
\end{figure}

In order to determine the value of $G_{e\nu}$, we assume that 
Eq.~(\ref{WNU}) describes the instantaneous emitted power 
from an electron 
with {\it arbitrary} world line at the point where it has proper 
acceleration $a$. This is indeed verified for photon (see Larmor formula
in Ref.~\cite{JAC}) and scalar particle~\cite{CHM2} emission from 
accelerated sources. [We emphasize that this equivalence is not 
fully (although it is approximately) verified for 
Eq.~(\ref{GNU}), which depends in general on the source's world line.]
Thus we will impose that Eq.~(\ref{WNU}) gives the radiated power
for the neutrino emission from circularly moving relativistic electrons
in a uniform magnetic field $B$ provided that 
$a=\gamma e B/m_e \ll m_e$ (no-recoil condition). 
Here $\gamma$ is the usual Lorentz factor for the electron and $e$ is its 
electric charge. The differential emission rate of $\nu_e \bar \nu _e$ pairs 
in a background magnetic field was calculated in detail~\cite{BKS} 
(see Ref.~\cite{LP} for the form used below):
\begin{eqnarray}
\frac{d\Gamma^{LP}_{\nu \bar \nu}}{ds}\;&=&\;
\frac{G_F^2 m_e^4}{16(2\pi)^3}
\frac{m_e}{\gamma}
\frac{\chi^5s^{3+1/2}}{(1+\chi s^{3/2})^4}
\nonumber \\
& &
\times
\left\{
\left(
C_V^2+C_A^2
\right)
\frac{\chi^2s^3}{(1+\chi s^{3/2})}
\int_s^\infty
\left[
2+\frac{1}{3}(2s+y)(y-s)^2
\right]
{\rm Ai}(y)\;dy\;
\right.
\nonumber \\
& &
+
\left(
C_V^2+C_A^2
\right)
\left[
\int_s^\infty
\left[ 
6+(y-s) 
\left(
s^2+(s-y)^2
\right)
\right]
{\rm Ai}(y)\;dy\;
-s{\rm Ai}(s)
\right]
\nonumber \\
& &
\left.
+
8sC_A^2 
\left[
\frac{3}{4}
\left(
\int_s^\infty
(s-y)^2
{\rm Ai}(y)\;dy
\right)
\;+
{\rm Ai}(s)
\right]
\right\}\;,
\label{GLP}
\end{eqnarray}
where $\chi\equiv a/m_e$, Ai($z$) is the Airy function, and
$s\in [0,\gamma /\chi]$ is defined such that 
\begin{equation}
\omega_\nu+\omega_{\bar \nu}\equiv
\frac{m_e\gamma \chi s^{3/2}}{(1+\chi s^{3/2})}\;.
\label{SOMEGA}
\end{equation}
The parameters $C_V$ and $C_A$ give the vector and axial 
contributions to the
electric current, respectively.
Using Eqs.~(\ref{GLP}) and (\ref{SOMEGA})
we have, in the limit $\chi \ll 1$,
\begin{eqnarray}
{\cal W}^{LP}_{\nu {\bar \nu}} &=&
\int_0^{\gamma/\chi}ds\;
\left(
\omega_\nu+\omega_{\bar \nu}
\right)
\frac{d\Gamma^{LP}_{\nu{\bar \nu}}}{ds}
\nonumber \\
&=&
\frac{5\;(2\;C_V^2+23\;C_A^2)}{108\pi^3}\;
G_F^2\;m_e^6\chi^6\;.
\label{W2}
\end{eqnarray} 
Letting $C_V^2=0.93$ and $C_A^2=0.25$ \cite{KLY}, we have
$
{\cal W}^{LP}_{\nu {\bar \nu}} =
1.14\times 10^{-2}\;G_F^2\;a^6\; .
$
By comparing this expression with our Eq.~(\ref{WNU}) we obtain
$
G_{e\nu} = 2.45 \;G_F.
$
In Figs.~\ref{fig4} and \ref{fig5} we plot Eqs.~(\ref{GNU}) and (\ref{WNU}), 
respectively, for uniformly accelerated electrons with $a\leq m_e$.

\begin{figure}
\begin{center}
\mbox{\epsfig{file=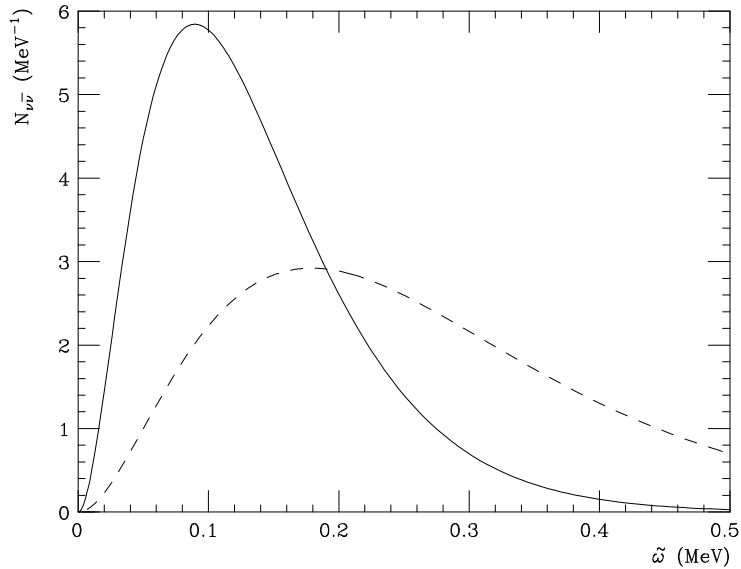,width=0.6\textwidth,angle=90}}
\end{center}
\vskip -1 cm
\caption{ The normalized energy distribution of the emitted neutrinos 
(and antineutrinos) is plotted for two values of the 
electron's proper acceleration: $a=0.1$ MeV (full line) and 
$0.2$ MeV (dashed line). Note that the typical energy
of the emitted particles 
(in the inertial frame instantaneously
at rest with the electron)
is given by ${\tilde \omega} \approx a$. }
\label{fig6}
\end{figure}

The normalized energy distribution of emitted neutrino-antineutrino
\begin{equation}
{\cal N}_{\nu {\bar \nu}}
\equiv
\frac{1}{\Gamma_{\nu {\bar \nu}}}
\frac{d \Gamma_{\nu {\bar \nu}}}{d{\tilde \omega}_\nu}\;
\label{Nnu}
\end{equation}
is plotted in Fig.~\ref{fig6} for electrons with proper acceleration
$a=0.1$ MeV and $0.2$ MeV, where [see Eq.~(\ref{dG5})]
\begin{equation}
\frac{d \Gamma_{\nu {\bar \nu}}}{d{\tilde \omega}_\nu}
\;=\;
\frac{2 G_{e\nu}^2}{\pi ^4 a}\;
{\tilde \omega}_\nu^2
\int_0^\infty
d{\tilde \omega}_{\bar \nu}
\;
{\tilde \omega}_{\bar \nu}^2 \;
K_0
\left[
2\left(
{\tilde \omega}_\nu+{\tilde \omega}_{\bar \nu}
\right)
/a
\right]\;.
\label{dGdw}
\end{equation}
(Neutrinos and antineutrinos have identical emission energy distribution.)
Note again that $a$ defines the typical energy of the emitted neutrinos.

\setcounter{equation}{0}


\section{Discussions}
\label{Discussions}

We have investigated the weak interaction emission of spin-$1/2$
fermions from classical and semiclassical currents.
As a particular application of modeling the accelerated particle by a 
semiclassical current, we have analyzed the inverse $\beta$ decay of 
uniformly accelerated protons.
We have shown that although protons in laboratory storage rings
are not likely to decay in this way, under some astrophysical conditions
high-energy protons in background magnetic fields may have a 
considerably short lifetime. Moreover, we have analyzed the modification
of the usual $\beta$ decay for uniformly accelerated neutrons.
This may be of some relevance when neutrons are
under the influence of strong background gravitational fields.
Although a full curved spacetime calculation is desirable to treat
these situations, our calculation should be a good approximation 
when the gravitational field is ``moderate''~\cite{CHM2}. In this case, 
neutrons  can be treated  as being accelerated in Minkowski space.

By restricting our semiclassical current to behave classically,
we were able to use our formalism to investigate the 
neutrino-antineutrino pair emission from uniformly accelerated 
electrons and compare our results with the 
ones in the literature obtained by quantizing the
electron field in a background magnetic field. Our formalism
allows the utilization of currents associated with more general 
world lines. Depending on the accuracy level required, however, 
one can use directly the formulas derived for uniformly 
accelerated currents. This may be particularly useful in some 
astrophysical situations.

\begin{flushleft}
{\bf{\large Acknowledgements}}
\end{flushleft}

The authors are thankful to J.C. Montero, V. Pleitez and A.A. Natale
for discussions. D.V. was fully supported by
Funda\c c\~ao de Amparo \`a Pesquisa do Estado de S\~ao Paulo
while G.M. was  partially supported by
Conselho Nacional de Desenvolvimento Cient\'\i fico e 
Tecnol\'ogico.

\appendix
\section*{Integration of 
          Eqs.~(\protect\ref{GAIII3}\protect)-(\protect\ref{WIII3}\protect)}

In order to solve Eqs.~(\ref{GAIII3}) and (\ref{WIII3})
let us consider the integral
\begin{equation}
{\cal I}_n^+ \;\equiv \;
\int_{0}^{\infty}dw\;
\frac{w^{n+2i\Delta M/a}}
{(w^2-1+2i\epsilon w)^{n+1}}\;,\;\;n\in{\rm N}\;.
\label{AP1}
\end{equation}
Note that the analytic extension of the integrand above has 
poles of order $(n+1)$ at $ w^{\pm} = \pm 1 
-i\epsilon +{\cal O}(\epsilon^2)$ . 
This implies that we can make $\epsilon =0$ in
Eq.~(\ref{AP1})
provided we contour the pole at $w^+=1$ by the upper half-plane,
i.e.:
\begin{equation}
{\cal I}_n^+ \; =\;
\int_{\gamma_+}dw\;
\frac{w^{n+2i\Delta M/a}}
{(w^2-1)^{n+1}}\;,
\label{IN+2}
\end{equation}
where $\gamma_\pm\equiv [0,1-\epsilon ']
\cup
\{1\pm \epsilon ' e^{i\theta}\;;\;
\theta \in [0,\pi]\}
\cup 
[1+\epsilon ',\infty)$ with $\epsilon ' \to 0_+$.
Using the residue theorem we see that
\begin{equation}
{\cal I}_n^- - {\cal I}_n^+\;=\;2\pi i\; {\rm Res}(f_n)_{w=1}\;,
\label{RES}
\end{equation}
where ${\cal I}_n^-$ is obtained substituting $\gamma_+$ by $\gamma_-$ in
Eq.~(\ref{IN+2}), and we  denote the residue value of the function
\begin{equation}
f_n(w)\equiv
\frac{w^{n+2i\Delta M/a}}
{(w^2-1)^{n+1}}\;
\label{FUNCTION}
\end{equation}
at the point $w=w^\pm$ by ${\rm Res}(f_n)_{w=w^\pm}$.
Now, let us define 
\begin{equation}
{\cal I}_n  \equiv 
\int_{-\infty}^{+\infty} dw\;
\frac{w^{n+2i\Delta M/a}}
{(w^2-1+2i\epsilon w)^{n+1}}\; ,
\end{equation}
which can be written for $\epsilon \to 0$ as
\begin{equation}
{\cal I}_n\;=\;(-1)^n e^{-2\pi\Delta M/a}\;{\cal I}_n^- +
{\cal I}_n^+\;.
\label{IN2}
\end{equation}
Since the integrand of ${\cal I}_n$ is analytic 
in the upper half-plane and goes to zero like $|w|^{-(n+2)}$ as 
$|w|\to \infty$, it follows that ${\cal I}_n=0$. As a consequence
Eqs.~(\ref{IN2}) and (\ref{RES}) imply
\begin{equation}
{\cal I}_n^+\;=\;
\frac{-2\pi i \; {\rm Res}(f_n)_{w=1}}{1+(-1)^ne^{2\pi\Delta M/a}}\;,
\label{I+3}
\end{equation}
with
\begin{equation}
{\rm Res}(f_n)_{w=w^\pm}\;=\;\frac{1}{n!}
\left.
\frac{d^n}{dw^n}\left\{(w-w^\pm)^{n+1}f_n(w)\right\}
\right| _{w=w^\pm}\;.
\label{RESULT}
\end{equation}
Using  function (\ref{FUNCTION}) to explicitly evaluate
Eq.~(\ref{RESULT}) for $n=5, 6$, we obtain
\begin{equation}
{\cal I}_5^+\;=\;
\frac{-\pi}{120\;a^5}
\left(
\frac{4a^4 \Delta M +5 a^2 \Delta M ^3 + \Delta M ^5}
{ e^{2\pi\Delta M/a}-1 }
\right)\;,
\label{I+5}
\end{equation}
and
\begin{equation}
{\cal I}_6^+\;=\;
\frac{i\;\pi}{46080\;a^6}
\left(
\frac{225\;a^6 +1036 \;a^4 \Delta M ^2 + 560\;a^2\Delta M ^4+
64\; \Delta M^6}
{e^{2\pi\Delta M/a}+1}
\right)
\;.
\label{I+6}
\end{equation}

\end{document}